\documentclass[prc,twocolumn,showpacs]{revtex4}
\usepackage{graphicx}
\begin{document}

\title{Critical temperature for $\alpha$-particle condensation
in asymmetric nuclear matter.}

\author{Takaaki Sogo, Gerd R\"opke}
\affiliation{Institut f\"ur Physik, Universit\"at Rostock, 
             D-18051 Rostock, Germany}
\author{Peter Schuck}
\affiliation{
Institut de Physique Nucl\'eaire, CNRS, UMR 8608, Orsay F-91406, France}

\affiliation{Universit\'e Paris-Sud, Orsay F-91505, France}

\affiliation{
Laboratoire de Physique et Mod\'elisation 
des Milieux Condens\'es, CNRS and Universit\'e Joseph Fourier, 
25 Avenue des Martyrs, Bo\^ite Postale 166, F-38042 Grenoble Cedex 9, France}

\begin{abstract}
The critical temperature for $\alpha$-particle condensation
in nuclear matter with  Fermi surface imbalance 
between protons and neutrons is determined.
The in-medium four-body Schr\"odinger equation, generalizing  
the Thouless criterion of the BCS transition, is applied using a 
Hartree-Fock wave function for the quartet projected
onto  zero total momentum in  matter with different 
chemical potentials for protons and neutrons.
\end{abstract}


\maketitle

\section{introduction}

Clustering
in nuclei and nuclear matter has become an increasingly studied subject 
recently~\cite{ths01,oh04,oyn04,sms06,ofe06,fhr07a,
fhr07b,freer07,chernych07,wakasa07,fyh08,fhr08,o08,slr09}. 
Since the $\alpha$-particle is the smallest doubly magic nucleus, it is 
thus very stable and excited heavier nuclei may exist in  weakly bound states 
consisting of nucleons, pairs, $\alpha$-particles, 
etc.~\cite{fhr07a,fhr07b,fhr08}. 
Recently, excited states of $^{12}$C and $^{16}$O, 
like the Hoyle state in $^{12}$C,
have been described as so-called ''$\alpha$ condensation'' states,
which can be pictured by  three or four $\alpha$-particles occupying 
the lowest $s$-state in an effective quartet mean-field 
potential~\cite{ths01,fyh08}.

On the other hand, 
$\alpha$-clustering in nuclear matter
has also been investigated as a possible phase in compact stars 
or as a precursor of  
$\alpha$-clustering in finite nuclei. 
The critical temperature for the $\alpha$ condensation 
has previously been calculated with the momentum projected 
Hartree-Fock approximation,
yielding results consistent with the one 
by the Faddeev-Yakubovsky method~\cite{slr09}.
The critical temperature is
an important quantity with respect to stellar nucleosynthesis
and star formation~\cite{st83,baldo99}.
However, in comparison with our previous studies,
bulk nuclear matter, like the one in neutron stars and other compact objects, 
is imbalanced with respect to the particle numbers of 
the protons and neutrons, implying different chemical potentials.
In the present paper, we investigate the critical temperature in
asymmetric nuclear matter.

The applied method is the same as in our previous work of Ref.~\cite{slr09},
where the in-medium four-body Schr\"odinger equation is computed 
employing Hartree-Fock wave functions under the constraint of 
total momentum equal to zero.
While we used four equal single particle wave functions
in the previous paper for symmetric nuclear matter, we shall employ
different wave functions 
for protons and neutrons in the present case because of the 
population imbalance with the different chemical potentials 
for protons and neutrons.

In the next section, we shortly will describe the formulation which we 
use to compute the 
critical temperature. In Section~\ref{sec-numerical}, 
we show the numerical results
for the critical temperature as a function of density and chemical potential
for various population ratios of protons versus neutrons.
Besides, we will display and discuss the quartet wave function at the critical 
temperature. Finally in Section~\ref{sec-summary},  we summarize.

\section{formulation}

For the asymmetric nuclear matter case, we will closely follow 
the formulation given in our previous paper~\cite{slr09} 
for  symmetric nuclear matter.

The Hamiltonian  is represented by
\begin{equation}
H
=H_0+V
=\sum_{1}\varepsilon_{1}c_{1}^\dagger c_{1}
+\frac{1}{4}\sum_{1234}\bar v_{12,34}c_{1}^\dagger c_{1}^\dagger c_{4} c_{3},
\end{equation}
where indices represent  momentum, spin, and isospin,
$\bar v_{12,1'2'}$ is the antisymmetric two-body interaction matrix,
and 
$\varepsilon_{i}=\frac{k^2_i}{2m^*}$.
We formally introduce an effective mass $m^*$ to 
account approximately for the exchange term of the mean field, whereas the 
direct term is incorporated into the chemical potential. However, 
since in the following we consider mostly only systems at very low density 
we will disregard mean field effects and take for the mass the bare one.

The four-body Schr\"odinger equation in the medium
with the eigenvalue $E$
is given by (see \cite{drs98,rss98,slr09,srs10} for details)
\begin{eqnarray}
\varepsilon_{1234}\psi_{1234}
+
\sum_{1'2'3'4'}
V_{1234;1'2'3'4'}\psi_{1'2'3'4'}
=E\psi_{1234}
\label{eq-imfbse}
\end{eqnarray}
where $\varepsilon_{1234}= \sum_{i=1}^4 \varepsilon_i$ 
and $V_{1234;1'2'3'4'}$ is of the form 
\begin{eqnarray}
&&V_{1234;1'2'3'4'}
=(1-f_1-f_2)\frac{1}{2}\bar v_{12,1'2'}\delta_{33'}\delta_{44'}
\nonumber \\
&&+(1-f_1-f_3)\frac{1}{2}\bar v_{13,1'3'}\delta_{22'}\delta_{44'}
+\mbox{permutation},
\end{eqnarray}
where $f_i$ is the Fermi-Dirac distribution with different 
chemical potentials for protons and neutrons. Otherwise, formally the 
in-medium four-body equation is exactly the same as in the symmetric case.
The condition of the transition to condensation, 
known as the Thouless criterion, is satisfied with 
$E=2\mu_{p}+2\mu_{n}$
at $T\to T_c$~\cite{thouless}, where
$\mu_{p}$ ($\mu_{n}$) is the proton (neutron) chemical potential.

For simplicity, we consider also for  proton-neutron imbalanced nuclear 
matter a spin-isospin-independent two body interaction. 
We do not think that this is much less justified than 
in the symmetric case~\cite{slr09} where this was very 
successful, indeed.
We again employ for the 
four body wave function $\psi_{1234}$ 
the Hartree-Fock ansatz, projected on zero total momentum:
\begin{eqnarray}
\psi_{1234} 
&\to& 
\varphi_p(\vec k_1)\varphi_p(\vec k_2)\varphi_n(\vec k_3)\varphi_n(\vec k_4)
\chi_0
\nonumber \\
&\times&
(2\pi)^3\delta(\vec k_1+\vec k_2+\vec k_3+\vec k_4)
\label{eq-phfwf}
\end{eqnarray}
where $\varphi_{\tau}(\vec k_i)=\varphi_{\tau}(|\vec k_i|)$ is
the $s$-wave single particle wave functions for protons 
($\tau=p$) and neutrons ($\tau=n$), 
respectively.  $\chi_0$ is the spin-isospin singlet wave function.

Substituting the ansatz of Eq.~(\ref{eq-phfwf}) into 
Eq.~(\ref{eq-imfbse}), and integrating over superfluous variables,
we obtain
\begin{eqnarray}
\varphi_{\tau}(k)
=\frac{-3{\cal B}_{\tau}(k)}{{\cal A}_{\tau}(k)+3{\cal C}_{\tau}(k)},
\quad (\tau=p,n)
\label{eq-spwf}
\end{eqnarray}
where
\begin{eqnarray}
{\cal A}_p(k)
&=&
\int \frac{d^3k_2}{(2\pi)^3}\frac{d^3k_3}{(2\pi)^3}\frac{d^3k_4}{(2\pi)^3}
\nonumber \\
&&\times
\left(\frac{k^2}{2m}+\frac{k_2^2}{2m}+\frac{k_3^2}{2m}+\frac{k_4^2}{2m}
-2\mu_p-2\mu_n\right)
\nonumber \\
&&\times
(\varphi_p(\vec k_2))^2(\varphi_n(\vec k_3))^2(\varphi_n(\vec k_4))^2
\nonumber \\
&&\times
(2\pi)^3\delta(\vec k+\vec k_2+\vec k_3+\vec k_4),
\label{eq-aterm}\\
{\cal B}_p(k)
&=&
\int \frac{d^3k_2}{(2\pi)^3}\frac{d^3k_3}{(2\pi)^3}\frac{d^3k_4}{(2\pi)^3}
\frac{d^3k_1'}{(2\pi)^3}\frac{d^3k_2'}{(2\pi)^3}
\nonumber \\
&&\times
(1-f_p(k)-f_p(k_2))v_{\vec k\vec k_2,\vec k_1'\vec k_2'}
\nonumber \\
&&\times
(2\pi)^3\delta(\vec k+\vec k_2-\vec k_1'-\vec k_2')
\nonumber \\
&&\times
\varphi_p(\vec k_1')\varphi_p(\vec k_2)\varphi_p(\vec k_2')
(\varphi_n(\vec k_3))^2(\varphi_n(\vec k_4))^2
\nonumber \\
&&\times
(2\pi)^3\delta(\vec k+\vec k_2+\vec k_3+\vec k_4)
\nonumber \\
&+&2
\int \frac{d^3k_2}{(2\pi)^3}\frac{d^3k_3}{(2\pi)^3}\frac{d^3k_4}{(2\pi)^3}
\frac{d^3k_1'}{(2\pi)^3}\frac{d^3k_3'}{(2\pi)^3}
\nonumber \\
&&\times
(1-f_p(k)-f_n(k_3))v_{\vec k\vec k_3,\vec k_1'\vec k_3'}
\nonumber \\
&&\times
(2\pi)^3\delta(\vec k+\vec k_3-\vec k_1'-\vec k_3')
\nonumber \\
&&\times
\varphi_p(\vec k_1')(\varphi_p(\vec k_2))^2
\varphi_n(\vec k_3)\varphi_n(\vec k_3')(\varphi_n(\vec k_4))^2
\nonumber \\
&&\times
(2\pi)^3\delta(\vec k+\vec k_2+\vec k_3+\vec k_4),
\label{eq-bterm}\\
{\cal C}_p(k)
&=&
\int \frac{d^3k_2}{(2\pi)^3}\frac{d^3k_3}{(2\pi)^3}\frac{d^3k_4}{(2\pi)^3}
\frac{d^3k_3'}{(2\pi)^3}\frac{d^3k_4'}{(2\pi)^3}
\nonumber \\
&&\times
(1-f_n(k_3)-f_n(k_4))v_{\vec k_3\vec k_4,\vec k_3'\vec k_4'}
\nonumber \\
&&\times
(2\pi)^3\delta(\vec k_3+\vec k_4-\vec k_3'-\vec k_4')
\nonumber \\
&&\times
(\varphi_p(\vec k_2))^2\varphi_n(\vec k_3)\varphi_n(\vec k_3')
\varphi_n(\vec k_4)\varphi_n(\vec k_4')
\nonumber \\
&&\times
(2\pi)^3\delta(\vec k+\vec k_2+\vec k_3+\vec k_4)
\nonumber \\
&+&2
\int \frac{d^3k_2}{(2\pi)^3}\frac{d^3k_3}{(2\pi)^3}\frac{d^3k_4}{(2\pi)^3}
\frac{d^3k_2'}{(2\pi)^3}\frac{d^3k_3'}{(2\pi)^3}
\nonumber \\
&&\times
(1-f_p(k_2)-f_n(k_3))v_{\vec k_2\vec k_3,\vec k_2'\vec k_3'}
\nonumber \\
&&\times
(2\pi)^3\delta(\vec k_2+\vec k_3-\vec k_2'-\vec k_3')
\nonumber \\
&&\times
\varphi_p(\vec k_2)\varphi_p(\vec k_2')
\varphi_n(\vec k_3)\varphi_n(\vec k_3')(\varphi_n(\vec k_4))^2
\nonumber \\
&&\times
(2\pi)^3\delta(\vec k+\vec k_2+\vec k_3+\vec k_4)
\label{eq-cterm}
\end{eqnarray}
with the symmetric two-body vertex 
$v_{\vec k_2\vec k_3,\vec k_2'\vec k_3'}$. The corresponding expressions for 
the neutrons
${\cal A}_n$, ${\cal B}_n$, and ${\cal C}_n$
are obtained by exchanging indices  $p \leftrightarrow n$
for ${\cal A}_p$, ${\cal B}_p$, and ${\cal C}_p$
in Eqs.~(\ref{eq-aterm}), (\ref{eq-bterm}) and (\ref{eq-cterm}).
The Fermi distribution functions are 
\begin{eqnarray}
f_{\tau}(k)=\frac{1}{e^{(\frac{k^2}{2m}-\mu_{\tau})/T}+1},
\quad (\tau=p,n).
\label{eq-fddf}
\end{eqnarray}

Here, comparing with symmetric nuclear matter~\cite{slr09},
in the imbalanced nuclear matter case,  
two coupled equations 
are obtained for protons and neutrons.

\section{\label{sec-numerical}numerical calculation}

\begin{figure}
\includegraphics[width=74mm]{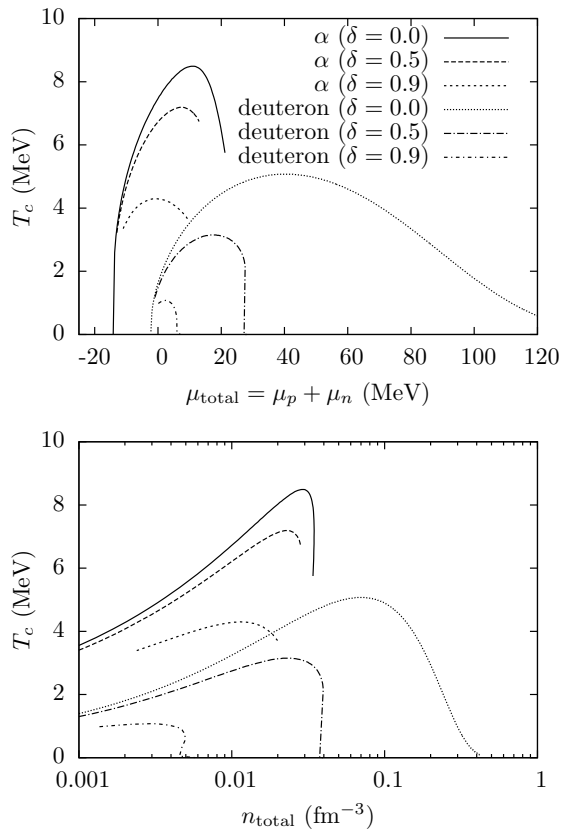}
\caption{\label{fig-ntotalvstc}
Critical temperature as functions of 
the total chemical potential $\mu_{\rm total}=\mu_p+\mu_n$ (top)
and the total free density $n_{\rm total}$ (bottom). 
The density ratio $\delta$
is in  Eq.~(\ref{eq-densityratio}).}
\end{figure}

As seen from Eqs.~(\ref{eq-aterm})-(\ref{eq-cterm}),
since the wave functions $\varphi_{p,n}(k)$ are mixed up in ${\cal A}_{p,n}$, 
${\cal B}_{p,n}$, and ${\cal C}_{p,n}$,
Eq.~(\ref{eq-spwf}) constitutes in fact two coupled non linear equations to be 
solved self-consistently by iteration.
The critical temperature is derived from  
the condition 
\begin{eqnarray}
&&
\int \frac{d^3k}{(2\pi)^3} \varphi_{p}(k)
[({\cal A}_{p}(k)+3{\cal C}_{p}(k))\varphi_{p}(k)+3{\cal B}_{p}(k)]
\nonumber \\
&&
=0,
\label{eq-condition1}
\\
&&
\int \frac{d^3k}{(2\pi)^3} \varphi_{n}(k)
[({\cal A}_{n}(k)+3{\cal C}_{n}(k))\varphi_{n}(k)+3{\cal B}_{n}(k)]
\nonumber \\
&&
=0.
\label{eq-condition2}
\end{eqnarray}
Given a set of chemical potentials $\mu_{p}$ and $\mu_{n}$
one can solve the two coupled equations 
(\ref{eq-condition1}) and (\ref{eq-condition2}) 
in adjusting the temperature, to be identified with the 
critical temperature, $T_c$, as the single parameter.

For the vertex $v_{\vec k_1\vec k_2,\vec k_1'\vec k_2'}$
in Eqs.~(\ref{eq-bterm}) and (\ref{eq-cterm}), we again take
the separable potential of~\cite{slr09}:
\begin{eqnarray}
v_{\vec k_1\vec k_2,\vec k_1'\vec k_2'}
&=&
\lambda e^{-\frac{(\vec k_1-\vec k_2)^2}{4b^2}}
e^{-\frac{(\vec k_1'-\vec k_2')^2}{4b^2}}
\nonumber \\
&\times&
(2\pi)^3\delta(\vec k_1+\vec k_2-\vec k_1'-\vec k_2').
\label{eq-separable}
\end{eqnarray}
The parameters $\lambda$ and $b$ are adjusted 
to the binding energy ($-28.3$MeV)  and to the
rms radius ($1.71$fm) of the isolated $\alpha$-particle;
$\lambda=-992$MeV fm$^3$ and $b=1.43$fm$^{-1}$. As already mentioned, in the 
symmetric case our procedure to solve the in-medium four-body equation with 
our ansatz (\ref{eq-phfwf}) and the separable interaction (\ref{eq-separable}) 
gave excellent agreement with a full Faddeev-Yakubovsky 
solution using 
the Malfliet-Tjon interaction (MT I-III)~\cite{slr09}.
We, therefore, think that our procedure is valid in the present case as well.

Fig.~\ref{fig-ntotalvstc} shows the critical temperature 
of $\alpha$ condensation as a function of the total chemical 
potential $\mu_{\rm total}=\mu_p+\mu_n$. 
We see that $T_c$ decreases as the asymmetry, given 
by the parameter
\begin{eqnarray}
\delta=\frac{n_n-n_p}{n_n+n_p},
\label{eq-densityratio}
\end{eqnarray}
increases. This is in analogy with the deuteron case (also shown) which 
already had been treated in~\cite{afr93,lns01}. 
On the other hand it also is interesting 
to show $T_c$ as a function of the free density 
which is
\begin{eqnarray}
n_{\rm total}&=&n_p+n_n 
\label{eq-totaldensity}\\
n_p&=&2\int \frac{d^3k}{(2\pi)^3}f_p(k) \\
n_n&=&2\int \frac{d^3k}{(2\pi)^3}f_n(k), 
\end{eqnarray}
where the factor two in front of the integral comes from the spin degeneracy,
and $f_{p,n}(k)$ is as in Eq.~(\ref{eq-fddf}).
It should be emphasized, however, that in the above relation between density 
and chemical potential, the free gas relation is used and correlations in the 
density have 
been neglected. In this sense the dependence of $T_c$ on density only is 
indicative, more valid at the higher density side. The very low density part 
where the correlations play a more important role shall be treated in a future 
publication. It should, however, be stressed that the dependence of $T_c$ on 
the chemical potential as in the upper panel if Fig.~\ref{fig-ntotalvstc}, 
stays unaltered.

The fact that for more asymmetric matter the transition temperature decreases, 
is natural, since as the Fermi levels become 
more and more unequal, the proton-neutron correlations will be suppressed.
For small $\delta$'s, i.e., close to the symmetric case, 
$\alpha$ condensation (quartetting) breaks 
down at smaller density
(smaller chemical potential) than 
deuteron condensation (pairing). 
This effect 
has already been discussed in our previous work for symmetric nuclear 
matter~\cite{slr09,rss98}.
For large $\delta$'s, i.e. strong asymmetries,  the behavior is opposite,
i.e., deuteron condensation breaks down at smaller densities than 
$\alpha$ condensation, 
because the small binding energy of the deuteron can not
compensate the difference of the chemical potentials.

More precisely,
for small $\delta$'s,
the deuteron with zero center of mass momentum
is only weakly influenced by 
the density or the total chemical potential 
as can seen in Fig.~\ref{fig-ntotalvstc}.
However, as $\delta$ increases, 
the different chemical potentials for protons and neutrons very much
hinders the formation of proton-neutron Cooper pairs in the isoscalar 
channel for rather obvious reasons.
The point to make here is that because of the much stronger binding per 
particle of the $\alpha$-particle,
the latter is much less influenced by the increasing difference 
of the chemical potentials.
For the strong asymmetry $\delta=0.9$ 
in Fig.~\ref{fig-ntotalvstc}
then finally $\alpha$-particle condensation can exist
up to $n_{\rm total}=0.02$fm$^{-3}$ ($\mu_{\rm total}=9.3$MeV),
while the deuteron condensation exists only up to 
$n_{\rm total}=0.005$fm$^{-3}$ ($\mu_{\rm total}=6.0$MeV).

\begin{figure*}
\includegraphics[width=170mm]{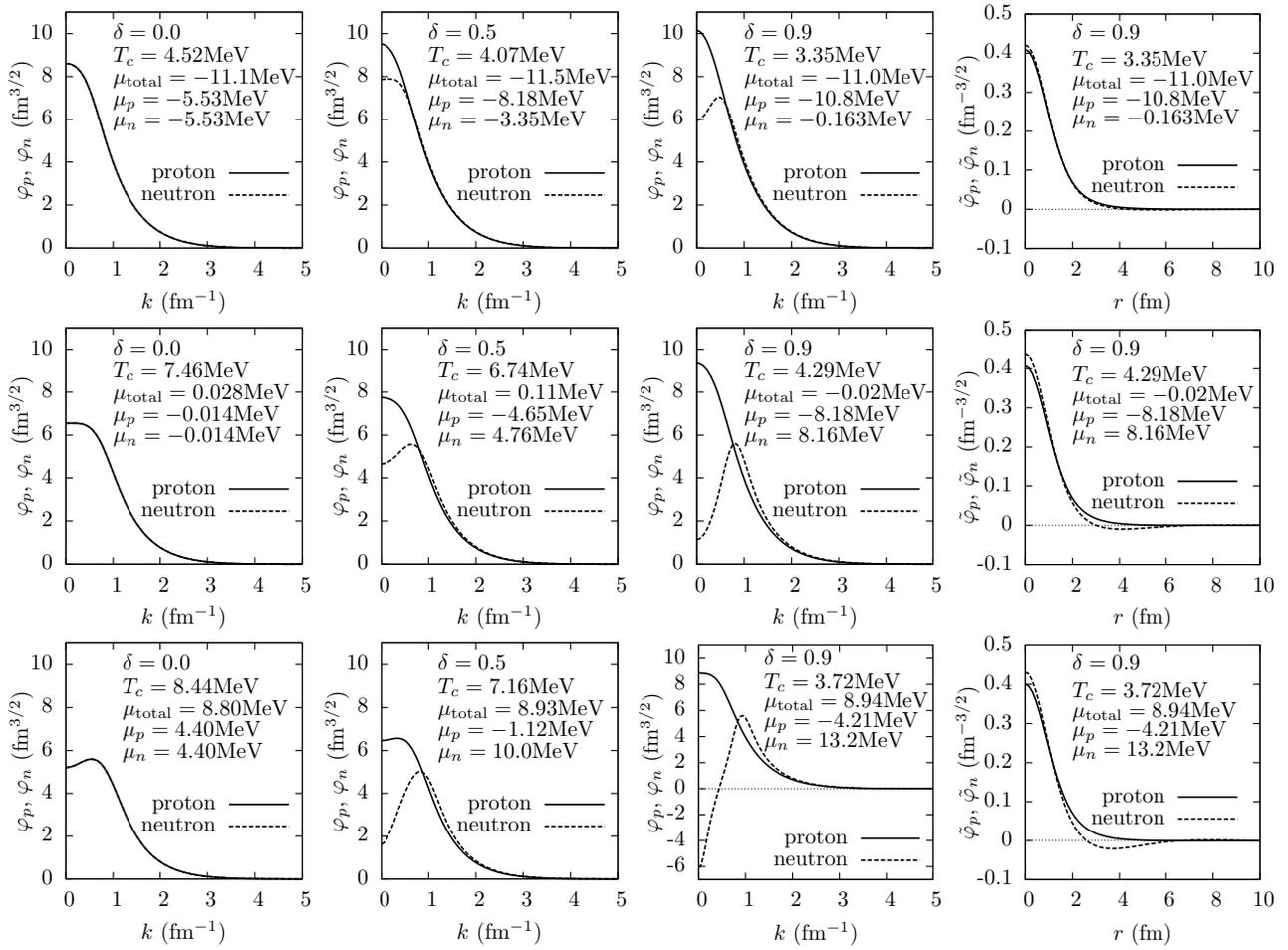}
\caption{\label{fig-spwf}
The momentum-space single particle wave functions $\varphi_p$, $\varphi_n$
at the critical temperature $T_c$
as function of $k$ for $\delta=0.0$, $0.5$ $0.9$,
and the real-space wave functions $\tilde \varphi_p$, $\tilde \varphi_n$
as a function of $r$ for $\delta=0.9$
derived from the Fourier transform of $\varphi_{p,n}(k)$ 
with $\tilde \varphi_{p,n}(r)=
\int d^3 ke^{i\vec k \cdot \vec r} \varphi_{p,n}(k)/(2\pi)^3$.
The top, middle and bottom figures are for 
$\mu_{\rm total}=\mu_p+\mu_n\sim-11$MeV, $\sim 0.0$MeV, and $\sim 9.0$MeV,
respectively.
The wave functions are normalized 
by $\int d^3 k \varphi_{p,n}^2(k)/(2\pi)^3=1$.}
\end{figure*}

Overall, the behavior of $T_c$ is more or less as can be expected. We should, 
however, remark that the critical temperature for $\alpha$-particle 
condensation stays quite high, even for the strongest asymmetry considered 
here, namely $\delta$ = 0.9. This may be of importance for the 
possibility of $\alpha$-particle condensation 
in neutron stars and supernovae explosions~\cite{sto98,ls91}.

We also show the single particle wave functions of protons and neutrons, 
entering the quartet wave function~(\ref{eq-phfwf}),  
for various ratios of Fermi surface imbalance and chemical potentials 
in Fig.~\ref{fig-spwf}.
In most cases of Fig.~\ref{fig-spwf}, 
the momentum-space wave functions with negative chemical potentials 
are monotonically decreasing whereas
the ones with positive chemical potentials
have a dip at $k=0$.
However, the momentum-space wave functions
also develop a dip at $k=0$ 
even at a negative chemical potential
as the asymmetry takes on stronger values.
This can be seen, e.g. for $\delta=0.5$, $\mu_{\rm total}=8.93$MeV,
and $\delta=0.9$, $\mu_{\rm total}=-11.0$MeV 
in Fig.~\ref{fig-spwf}.
Furthermore, the neutron wave function in $k$-space
with large positive chemical potential develops a node. 
This behavior is similar to the wave functions 
in Ref.~\cite{slr09}.
As shown in Fig.~\ref{fig-spwf},
the dissymmetry of proton and neutron wave functions increases
as $\delta$ increases. As a consequence, the critical temperature 
decreases, and the $\alpha$ condensation breaks down at a more dilute density,
see Fig.~\ref{fig-ntotalvstc}. We also present in most right figures
of Fig.~\ref{fig-spwf} the proton and neutron 
wave functions in real space. In spite of the sometimes strong dissymmetry in 
momentum space, the proton and neutron wave functions are relatively 
more similar to one another in $r$-space. 
The neutron wave function develops a node  
as the total chemical potential $\mu_{\rm total}=\mu_{p}+\mu_{n}$ increases,
but the negative values of the wave function remain rather 
moderate.

\section{\label{sec-summary}Summary}

We reported on the critical temperature for the $\alpha$-particle condensation
as a function of the density and chemical potential
in asymmetric nuclear matter.
The four-body wave function in the medium is calculated 
with Hartree-Fock wave functions  projected onto 
zero total momentum, a procedure which was already very successful in the 
symmetric case. Not unexpectedly the transition temperature decreases with 
increasing asymmetry. However, it was shown that $T_c$ stays relatively high 
for very strong asymmetries, a fact of importance in the 
astrophysical context. 
The single particle wave functions of proton and neutron were also shown 
and discussed. The neutron wave function in momentum space
develops a node for strong 
asymmetries and high densities, 
a fact familiar from ordinary pairing. It also was shown that 
asymmetry affects deuteron pairing more strongly than $\alpha$-particle 
condensation. Therefore, at high asymmetries, if at all, $\alpha$-particle 
condensation dominates over pairing at all possible densities.

The condition of Eqs.~(\ref{eq-condition1}) and (\ref{eq-condition2})
corresponds to calculating the critical temperature 
at the point where the $\alpha$-particles in nuclear matter 
dissociate into four free particles, 
i.e. at the Mott transition temperature. 
In the higher densities considered here, this temperature is consistent with 
the phase transition temperature between superfluid and normal phase.
In the strong coupling limit, this is not the case,
since the Bose Einstein condensation breaks down at lower temperature
than the one of dissociation of a bound state to free particles.
This problem in the BCS-BEC crossover can be solved 
by the Nozi\`eres and Schmitt-Rink theory, where 
the pair fluctuations are systematically included 
beyond the mean field approximation~\cite{ns85}.
The extension of this theory to the quartet condensation is difficult but 
will be attempted in future work. 

\acknowledgments

This work is supported by DFG No. RO905/29-2.


\begin{thebibliography}{99}
\bibitem{ths01}
A. Tohsaki, H. Horiuchi, P. Schuck, G. R\"opke, 
Phys. Rev. Lett. {\bf 87}, 192501 (2001).

\bibitem{oh04}
S. Ohkubo and Y. Hirabayashi, 
Phys. Rev. C {\bf 70}, 041602 (2004) 

\bibitem{oyn04}
H. Ohta, K. Yabana, and T. Nakatsukasa
Phys. Rev. C {\bf 70}, 014301 (2004). 

\bibitem{sms06}
A. Sedrakian, H. M\"uther, P. Schuck, Nucl Phys. {\bf A766}, 97 (2006). 

\bibitem{ofe06}
W. von Oertzen, M. Freer, and Y. Kanada-Enyo,
Phys. Reports {\bf 432}, 43 (2006).

\bibitem{fhr07a}
Y. Funaki, H. Horiuchi, G. R\"opke, P. Schuck, A. Tohsaki, T. Yamada, 
Nucl. Phys. News {\bf 17}, 11 (2007).

\bibitem{fhr07b}
Y. Funaki, H. Horiuchi, G. R\"opke, P. Schuck, A. Tohsaki, T. Yamada, 
Progress in Particle and Nuclear Physics {\bf 59}, 285 (2007).

\bibitem{freer07}
M. Freer, 
Rep. Prog. Phys. {\bf 70} 2149 (2007).

\bibitem{chernych07}
M. Chernykh, H. Feldmeier, T. Neff, P. von Neumann-Cosel, and A. Richter, 
Phys. Rev. Lett. {\bf 98}, 032501 (2007).

\bibitem{wakasa07}
T. Wakasa, E. Ihara, K. Fujita, Y. Funaki, K. Hatanaka, 
H. Horiuchi, M. Itoh, J. Kamiya, G. R\"opke, H. Sakaguchi, 
N. Sakamoto, Y. Sakemi, P. Schuck, Y. Shimizu, M. Takashina, 
S. Terashima, A. Tohsaki, M. Uchida, H.P. Yoshida, M. Yosoi, 
Phys. Lett. B {\bf 653}, 173 (2007).

\bibitem{fyh08}
Y. Funaki, T. Yamada, H. Horiuchi, G. R\"opke, P. Schuck, and A. Tohsaki,
Phys. Rev. Lett. {\bf 101}, 082502 (2008).

\bibitem{fhr08}
Y. Funaki, H. Horiuchi, G. R\"opke, P. Schuck, A. Tohsaki, and T. Yamada,
Phys. Rev. C {\bf 77}, 064312 (2008).

\bibitem{o08}
W. von Oertzen, {\it et. al.}, 
Eur. Phys. J. A {\bf 36}, 279 (2008). 


\bibitem{slr09}
T. Sogo, R. Lazauskas, G. R\"opke, and P. Schuck,
Phys. Rev. C {\bf 79}, 051301(R) (2009).

\bibitem{st83}
S. L. Shapiro and S. A. Teukolsky
{\it Black Holes, White Dwarfs and Neutron Stars: 
The Physics of Compact Objects},
(Wiley, New York, 1983). 

\bibitem{baldo99}
M. Baldo, 
{\it Nuclear Methods and the Nuclear Equation of State}, 
(World Scientific, Singapore, 1999).

\bibitem{rss98}
G. R\"opke, A. Schnell, P. Schuck, and P. Nozi\`eres,
Phys. Rev. Lett. {\bf 80}, 3177 (1998).

\bibitem{srs10}
T. Sogo, G. R\"opke, and P. Schuck,
Phys. Rev. C {\bf 81}, 064310 (2010).

\bibitem{drs98}
J. Dukelsky, G. R\"opke, and P. Schuck,
Nucl. Phys. {\bf A628}, 17 (1998).

\bibitem{thouless}
D. J. Thouless, Ann. Phys. {\bf 10}, 553 (1960).

\bibitem{afr93}
T. Alm, B. L. Friman, G. R\"opke, and H. Schulz,
Nucl. Phys. {\bf A551}, 45 (1993).

\bibitem{lns01}
U. Lombardo, P. Nozi\`eres, P. Schuck, H.-J. Schulze, 
A. Sedrakian. Phys. Rev. C {\bf 64}, 064314 (2001).

\bibitem{ls91}
J. M. Lattimer and F. D. Swesty,
Nucl. Phys. {\bf A535}, 331 (1991).

\bibitem{sto98}
M. Shen, H. Toki, K. Oyamatsu, and K. Sumiyoshi,
Prog. Theor. Phys. {\bf 100}, 1013 (1998).

\bibitem{ns85}
P. Nozi\`eres and S. Schmitt-Rink, 
J. Low Temp. Phys. {\bf 59}, 195 (1985).


\end{thebibliography}
\end{document}